\documentclass{PoS}
\usepackage{amsmath, shuffle,cite}
\title{{\footnotesize DESY 17-225, DO-TH 17/37 \hfill {\tt arXiv:yymmdd.xxxxx}}\\
Special functions, transcendentals and their numerics
\thanks{This work was supported in part 
by 
the
Austrian Science Fund (FWF) grant SFB F50 (F5006-N15, F5009-N15) and
the European Commission through contract PITN-GA-2012-316704 (HIGGSTOOLS).}}

\ShortTitle{Special functions, transcendentals and their numerics}

\author{Jakob Ablinger$^a$\footnote{jablinge@risc.uni-linz.ac.at}, Johannes 
Bl\"umlein$^b$\footnote{Johannes.Bluemlein@desy.de}, Mark 
Round$^{a,b}$\footnote{speaker,~mark.round@desy.de}, and Carsten 
Schneider$^a$\footnote{cschneid@risc.uni-linz.ac.at}
\\
$^a$~Research Institute for Symbolic Computation (RISC), Johannes Kepler University, \\ ~~~Altenberger 
Stra{\ss}e 69, 4040, Linz, Austria\\
$^b$~Deutsches Elektronen-Synchroton (DESY), Platanenalle 6, D-15738, Zeuthen, Germany}

\abstract{Cyclotomic polylogarithms are reviewed and new results concerning the special constants that 
occur are presented.  This also allows some comments on previous literature results using PSLQ.}

\FullConference{13th International Symposium on Radiative Corrections (Applications of Quantum Field Theory to Phenomenology)\\
         25-29 September, 2017 \\
         St. Gilgen, Austria}

\begin{document}

\section{Outline}

\vspace*{1mm}
\noindent
Through the usual procedure of introducing Feynman (or Schwinger) parameters one can perform the momentum 
integrals 
of Feynman diagrams.  This leads to multiply nested integrals over $(0,1)$.  In principal one should 
not expect such integrals to 
always be expressible in terms of elementary and known special functions.  Immediately then one must ask what it 
means to `solve' a Feynman diagram.  To compare theoretical results to experimental data a solution could be 
interpreted as a form which can be evaluated quickly.  On mathematical grounds, a solution could be interpreted as 
expressing integrals in terms of a set of basis objects which admit no further relations amongst them. 
Naturally one would 
like to achieve both goals simultaneously.  Perhaps the most basic set of objects, beyond elementary and known 
special functions, found in particle physics are the harmonic polylogarithms~\cite{Remiddi:1999ew} or, in Mellin 
space, the harmonic 
sums \cite{HSUM}.  The 
harmonic polylogarithms are a generalisation of the classical polylogarithm, see e.g. \cite{lewin1981} and the 
Nielsen 
integrals \cite{NIELSEN}.  These harmonic polylogarithms 
generalise to the cyclotomic polylogarithms~\cite{Ablinger:2011aa} and then to the generalised (or Goncharov) 
polylogarithms and beyond, see for example~\cite{Moch:2001zr,Vollinga:2004sn,Ablinger:2011aa,Ablinger:2013cf} for a 
literature.\footnote{For a recent survey on the different function spaces see \cite{Ablinger:2013eba}.}  
The 
questions are what relations exist between these polylogarithms and how can they be efficiently evaluated?

All of these types of polylogarithms are defined through iterated integrals\footnote{The notation 
used reflects that of the Mathematica package~\texttt{HarmonicSums}
\cite{RISC4542,HARMONICSUMS,Ablinger:2011te,Ablinger:2013cf,Ablinger:2014bra}.} 

\begin{align}
&H(\vec{a};x) := \int_0^x dt f_{a_1}(t) H(\vec{a}_{i>1};t) , && H(\emptyset;x):=1 .
\end{align}
Here the vector notation $\vec{a}$ is something of an abuse of notation as $\vec{a}$ will be frequently 
treated as an ordered list.  Meanwhile $\vec{a}_{i>1}$ refers to $\vec{a}$ with the first element 
removed.  
The components of $\vec{a}$ are known as indices and the length of $\vec{a}$ is known as the weight of the polylogarithm.  Making a choice for the functions $f_i$ one falls into the various classes of polylogarithms.  All of the classes used in particle physics include $f_i(t)=0$ which necessitates an additional definition for the weight $n$ polylogarithm whose indices are all zero, $H(\vec{0};x):=\tfrac{1}{n!}\log^n x$.  The treatment here will be largely formal and little further attention will be given to questions of existence.  

Iterated integrals have several properties on general 
grounds~\cite{chen1977,REUTENAUER,Blumlein:2003gb}.  There are 
identities 
from using integration by parts and from the shuffle product.   The shuffle of two lists $a$ and $b$ can be defined recursively as follows,
\begin{align}
&a \shuffle \emptyset :=\lbrace a\rbrace,
&&\emptyset \shuffle b := \lbrace b \rbrace,
&&a\shuffle b := a_1^\frown (a_{i>1}\shuffle b) \cup  b_1^\frown (a\shuffle b_{i>1}) .
\end{align}
Here $a_1^\frown X$ is the list of lists $X$ with $a_1$ pre-pended to \textit{each} element of $X$.  As defined here the result of $a\shuffle b$ will always be a list of lists.  The shuffle product is,
\begin{align}
H(\vec{a};x)H(\vec{b};x) = \sum_{\vec{c}\in\vec{a}\shuffle \vec{b}} H(\vec{c};x).
\end{align}
This turns the space of polylogarithms into a graded algebra with the grading given by the weight.  
(The so-called multiple polylogarithms also admit a stuffle product and associated stuffle identities 
which is omitted here for space, see e.g.~\cite{Ablinger:2013cf,hoffman1992, Vollinga:2004sn}.)  
By repeated integration by parts one can also obtain the following class of identities,
\begin{align}
H(\vec{a};x) = H(a_1;x)H(\vec{a}_{i>1};x) - H(a_2,a_1;x)H(\vec{a}_{i>2};x) + \ldots +(-1)^{n+1} H(a_n,a_{n-1},\ldots, a_1;x),
\end{align}
for a weight $n$ polylogarithm.

The relations of the generalized polylogarithms have been extensively studied in 
Ref.~\cite{Ablinger:2013cf}. Furthermore, it is conjectured that the symbol~\cite{goncharov2002, 
Duhr:2012fh, Duhr:2011zq} provides 
all relations amongst the generalised polylogarithms.  These relations can be used to express physical results 
numerically more simply however the topic is not yet exhausted.  It has been suggested that cyclotomic 
polylogarithm constants may obey additional relations beyond the shuffle, stuffle and integration by parts identities see~\cite{zhao2007} for a proper description of the idea.

There are also physics motivations for further study, most notably numerical implementations and PSLQ 
searches \cite{PSLQ}
.  To calculate the numerical value of a cross-section one must evaluate large numbers of 
polylogarithms which requires efficient numerical implementation.  This in turn also leads to the 
question of polylogarithms evaluated at 1.  This will be explained in Section~\ref{sec:cyclo} when the 
problem examined here is set-up.  Secondly, there has been progress in computing arbitrary precision 
values for amplitudes and using a PSLQ search to fit numerical factors.  Such work is helped by further 
knowledge about the special constants that occur in particle physics.

In this manuscript cyclotomic constants evaluated at 1 are studied.  It is explained how up to weight 2 
all PSLQ relations can be derived analytically using standard results about classical polylogarithms, which
is one of the known results of Ref.~\cite{Ablinger:2011aa}. 
This complements work already present 
in the literature~\cite{
Broadhurst:1998rz,
Kalmykov:2010xv,
Ablinger:2011aa,
Henn:2015sem} and itself is a part of a larger project~\cite{inprep}.

The remainder of the manuscript is organised as follows.  In the next section the cyclotomic 
polylogarithm class is set-up then the set of constants studied are introduced.  In 
Section~\ref{sec:pslq} it is explained how these relations can be calculated and the resulting special 
numbers are listed.  A short summary closes the manuscript.

\section{Cyclotomic polylogarithms\label{sec:cyclo}}

\vspace*{1mm}
\noindent
The $n^{\textrm{th}}$ cyclotomic polynomial $\Phi_n(x)$ may be defined in terms of the primitive roots of unity,
\begin{align}\label{eq:cyclodef}
\Phi_n(x): =\prod_{\substack{1\leq k \leq n \\ \gcd(k,n)=1}} \left(x- e^{2\pi i \tfrac{k}{n}}\right) .
\end{align}
The letters needed to define cyclotomic polylogarithms 
\cite{Ablinger:2011aa} are
\begin{equation}
f_{(k,l)} =  \frac{x^l}{\Phi_k(x)} ,
\end{equation}
so that now the indices of a cyclotomic polylogarithm have two components.  The particular values $f_{(1,0)}$, $f_{(2,0)}$ and $f_{(0,-1)}$ reproduce the usual harmonic polylogarithms (up to sign conventions).

Although the cyclotomic polynomials are defined in terms of roots of unity they give rise to real functions for $x\in(0,1)$ and therefore represent an elegant language for expressing physical results.  Alternatively, one may forego the advantages of a real representation, and factorise the cyclotomic polynomials leading to complex indices and the (larger class of) generalised polylogarithms.  In doing so one gains additional properties 
such as the scaling property and H\"older convolution which can be useful~\cite{Frellesvig:2016ske}.  

For the physical region, $x\in(0,1)$ a Maclaurin series can be used for numerics,
\begin{align}
\frac{1}{x-a} = -\frac{1}{a} \sum_{r=0}^\infty \left(\frac{x}{a}\right)^r ,
\end{align}
because by~\eqref{eq:cyclodef} $\left\vert\tfrac{x}{a}\right\vert <1$ for cyclotomic polylogarithms.  After inserting this into the iterated integral expression for a cyclotomic polylogarithm and performing all integrals by expanding everywhere one obtains a series representation.  However these series diverge as $x\rightarrow 1$ and therefore have poor convergence for large $x<1$.  The solution is to perform a particular variable transformation which expresses a polylogarithm with larger argument $x<1$ in terms of constants and polylogarithms with small argument $x>0$.  The formulae are a little lengthy so here a simple example transformation will suffice,
\begin{align}\label{eq:newrep}
H\left((1,0),(2,0);\frac{1-t}{1+t}\right) &= 
-H((2,0); 1)H((2,0); t) - H((2,0);1)H(0; t) + 2 H((2,0), (2,0); 1) \nonumber\\
&+ 
 H((2,0), (2,0); t) - H((2,0), (1,0); 1) + H((0,-1), (2,0); t) .
\end{align}
If $x=\tfrac{1-t}{1+t}$ then the region $x\in (\sqrt{2}-1,1)$ is mapped to $t\in (0,\sqrt{2}-1)$ thus a cyclotomic polylogarithm at a large argument can be expressed in terms of cyclotomic polylogarithms with small arguments that have faster converging series.  One can further 
exploit a so-called Bernoulli speed-up variable which writes the expansion in terms of logarithms of 
$x$ and gives faster convergence, see~e.g.~\cite{Gehrmann:2001pz}.  A consequence of this variable 
transform is the appearance of several cyclotomic polylogarithms evaluated at 1.  Here these are well-known 
special values,
\begin{align}
&H((2,0); 1) = \log 2 , &&H((2,0), (2,0); 1) = \frac{1}{2}\log^22 , && H((2,0), (1,0); 1) = -\textrm{Li}_2\left(\tfrac{1}{2}\right) ,
\end{align}
using the dilogarithm, $\textrm{Li}_2$.  This new representation~\eqref{eq:newrep}, in terms of the 
transformed argument $t$, is suitable for efficient numerical representation.  For harmonic 
polylogarithms this does not introduce any significant problems however in the cyclotomic case the 
constants generated are less well-studied.  Indeed at higher weights there are many new special numbers 
currently given little attention in the literature. 

In producing numerical representations for the cyclotomy $k\leq 6$ cyclotomic polylogarithms a large set of cyclotomic constants evaluated at 1 were generated.  It was therefore natural to study these numbers and their relations as part of the larger ongoing work of understanding polylogarithms.  After eliminating shuffle, stuffle and integration by parts identities there are still many prospective constants.  
By generating a large number of decimal places of precision for each constant one can 
try to build a basis by eliminating all PSLQ relations.  A similar exercise was conducted 
in~\cite{Henn:2015sem}.  An important difference is that here the motivation is the constants required 
for a numerical implementation and this led to the introduction of cyclotomy 12 objects as part of 
eliminating relations amongst the constants.  See~\cite{inprep} for a fuller explanation.  Thus the 
constants studied here include those from the previous studies such as~\cite{Broadhurst:1998rz,
Kalmykov:2010xv,
Ablinger:2011aa,
Henn:2015sem 
}.  Our concern is whether the PSLQ relations can be proven using known properties of polylogarithms.

\section{PSLQ Relations\label{sec:pslq}}

\vspace*{1mm}
\noindent
Weight 1 cyclotomic polylogarithms can be expressed in terms of logarithms and $\pi$.  
Combined with the Lindemann-Weierstra{\ss} theorem all relations amongst the 
cyclotomic polylogarithms can be proven. A small example illustrates the general case,
\begin{align}
H((12,3);1) = \int_0^1dt \frac{t^3}{1 - t^2 + t^4} = \frac{\pi}{6\sqrt{3}} .
\end{align}
Due to the introduction of cyclotomy 12 --- or equivalently the $12^{\textrm{th}}$ root of unity if 
preferred --- additional numbers beyond those in 
work~\cite{Broadhurst:1998rz,Kalmykov:2010xv,Henn:2015sem} are found.   In Ref.~\cite{Ablinger:2011aa} all weight 1 and weight 2
relations have been derived in analytic form.

At weight 2 the direct integrals are fully expressible in terms of logarithms and dilogarithms 
\cite{Ablinger:2011aa}. Using relations such as those in~\cite{lewin1981} all weight 2 PSLQ relations can be 
proven.  The first step is direct integration of the polylogarithm which can even be done using many 
computer algebra systems.  Alternatively one may use a well-known result that any weight 2 generalised 
polylogarithm can be expressed in terms of the dilogarithm and logarithm, see for 
example~\cite{Frellesvig:2016ske, Duhr:2011zq}.  Then by factoring cyclotomic polylogarithms into 
generalised polylogarithms the result follows.  The resulting logarithms do not represent further 
challenges.  However the dilogarithms present have complex arguments which can be decomposed into real 
and imaginary parts using a result by Kummer, as described in Refs.~\cite{Kirillov:1994en, 
lewin1981}, \begin{align}
\textrm{Li}_2(re^{i\theta}) =\textrm{Li}_2(r,\theta) +i\left[ w\log r +\frac{1}{2}\textrm{Cl}_2(2w)+\frac{1}{2}\textrm{Cl}_2(2\theta)-\frac{1}{2}\textrm{Cl}_2(2\theta+2w) \right] .
\end{align}
Here $w:=\arctan\tfrac{r\sin \theta}{1-r\cos\theta}$ and $\textrm{Cl}_2$ is Clausen's function.  
Additionally~\cite{lewin1981} lists many known particular values for the real part, 
$\textrm{Li}_2(r,\theta)$.  In order to complete the computation and remove all relations it 
was also required to use two identities due to Ramanujan~\cite{BERNDT},\footnote{These constants also
occur in the calculation of massive 3-loop operator matrix elements
\cite{Ablinger:2014yaa,Ablinger:2015tua}.}
\begin{eqnarray}
\textrm{Li}_2\left(-\tfrac{1}{3} \right) - \frac{1}{3}\textrm{Li}_2\left(\tfrac{1}{9} \right) &=& 
\frac{1}{6}\log^23-\frac{\pi^2}{18} ,\\
\textrm{Li}_2\left(\tfrac{1}{4} \right) + \frac{1}{3}\textrm{Li}_2\left(\tfrac{1}{9} \right) 
&=& \frac{\pi^2}{18} +2\log 2 \log 3- 2 \log^2 2-\frac{2}{3}\log^2 3 .
\end{eqnarray}
After using all these relations it was possible to arrive at the set of special numbers,
\begin{align*}
&\pi , 
&& \log 2,
&&\log 3,
&&\log(\sqrt{3}-1) ,\\
&\log(2-\sqrt{3}),
&&\textrm{Cl}_2\left(\tfrac{\pi}{3}\right),
&&\textrm{Cl}_2\left(\tfrac{\pi}{6}\right),
&&\textrm{Li}_2\left(\tfrac{1}{4}\right), \\
&\textrm{Li}_2(\tfrac{1}{2} (\sqrt{3}-1)),
&&\textrm{Li}_2(\tfrac{1}{4} (3 \sqrt{3}-5)),
&&\textrm{Li}_2(2- \sqrt{3}),
&&\textrm{Li}_2(4 \sqrt{3}-7).
\end{align*}
This set includes those numbers found in~\cite{Broadhurst:1998rz} plus additional numbers from the 
cyclotomy $12$ feature particular to this study. In our analysis we used the packages {\tt 
HarmonicSums} 
\cite{RISC4542,HARMONICSUMS,Ablinger:2011te,Ablinger:2013cf,Ablinger:2014bra} and {\tt Sigma} 
\cite{SIG1,SIG2} to eliminate known relations.

As a corollary of this work it was also possible 
to prove all the PSLQ relations at weights 1 and 2 conjectured in~\cite{Henn:2015sem}. In a different, 
but equivalent, basis representation the corresponding result has been given in \cite{Ablinger:2011aa}
in 2011 already.

\section{Summary}

\vspace*{1mm}
\noindent
Although much is known about polylogarithms and their special values at special arguments $x \in 
\mathbb{C}$ there is no complete theory yet and it has been suggested 
more may be available to learn.  Based on this long term goal a set of specific constants arising 
from a numerical implementation of cyclotomic polylogarithms were studied. It was possible to show 
that all relations amongst these constants are already present in the literature and a PSLQ search 
does not provide any new information for $w \leq 6$ for cyclotomy $k=6$. However, new 
relations were found for cyclotomy $k=12$. Analytic proofs of part of the relations have been 
obtained. More details will be presented in 
Ref.~\cite{inprep}.\footnote{After completion of this paper 
\cite{Schnetz:2017bko} appeared, in which aspects of bases in the cyclotomic case are discussed under 
specific conditions.}

 
\end{document}